\documentclass[aps,prd,amsmath,amssymb,longbibliography,
nofootinbib,twocolumn,showpacs,preprintnumbers]{revtex4-1}
\pdfoutput=1

\usepackage[toc,page]{appendix}
\usepackage{graphics}      
\usepackage{graphicx}     
\usepackage{epsfig}
\usepackage{epstopdf}
\usepackage{caption} 
\usepackage{subcaption} 
\usepackage{float} 

\usepackage[table]{xcolor}    
\usepackage{url}           
\usepackage{bm}            
\usepackage{mathrsfs}
\usepackage{breakurl}

\usepackage{hyperref}
\usepackage{color}
\usepackage{soul}
\usepackage{tikz}
\usetikzlibrary{calc}

\usepackage{pgfplots}
\pgfplotsset{width=5.7cm,compat=1.9}
\pgfplotsset{yticklabel style={text width=2.5em,align=right}}
\pgfplotsset{every axis/.append style={thick}}

\newcommand\be{\begin{equation}}
\newcommand\ee{\end{equation}}
\newcommand\nono{\nonumber}
\newcommand\bse{\begin{subequations}}
\newcommand\ese{\end{subequations}}
\newcommand\bea{\begin{eqnarray}}
\newcommand\eea{\end{eqnarray}}

\newcommand{\pd}{\partial}

\newcommand\ringring[1]{%
  {
   \mathop{\kern0pt #1}\limits^{
     \vbox to-1.85ex{
       \kern-2ex 
       \hbox to 0pt{\hss\normalfont\kern.1em \r{}\kern-.45em \r{}\hss}%
       \vss 
     }
   }
  }
}
\newcommand{\q}{\quad}
\newcommand{\qq}{\qquad}

\allowdisplaybreaks
\begin{document}

\title{Extended body dynamics in general relativity: Hyperelastic models}

\author{Nishita Jadoo}
\affiliation{Department of Physics, North Carolina State University, Raleigh, NC 27695}
\author{J. David Brown}
\affiliation{Department of Physics, North Carolina State University, Raleigh, NC 27695}
\author{Charles R. Evans}
\affiliation{Department of Physics and Astronomy, University of North Carolina, Chapel Hill, NC 27599}

\begin{abstract}
We present a numerical framework for modeling extended hyperelastic bodies based on a Lagrangian formulation of general relativistic elasticity theory. We use finite element methods to discretize the body, then use the semi--discrete action to derive ordinary differential equations of motion for the discrete nodes. The nodes are evolved in time using fourth--order Runge--Kutta. We validate our code against the normal modes of oscillation of a hyperelastic sphere, which are known analytically in the limit of small (linear), slow (Newtonian) oscillations. The algorithm displays second order convergence. This numerical framework can be used to 
obtain the orbital motion and internal dynamics of a hyperelastic body of any shape, for any spacetime metric, and for varying hyperelastic energy models. 
\end{abstract}

\maketitle

\section{Introduction}

The problem of motion in general relativity has a long history. Einstein was interested in whether the laws of motion of material points can be derived from the vacuum field equations. With Grommer, \cite{einstein1927} he showed that if a point particle is treated as a singularity in spacetime, it follows a geodesic. This gives the motion of point particles but not extended bodies. The first person to describe an extended body in general relativity was Mathisson \cite{Mathisson1937}. Mathisson defines a multipole expansion of the body using the body's stress--energy--momentum (SEM) tensor with the single pole (monopole) defining the mass and the dipole and quadrupole defining the ``rotation moment''. Subsequently, many others have worked on this problem following a similar method, including Papapetrou \cite{Papapetrou1951} who gives the equations of motion of spinning particles to dipole order. See also Refs.~\cite{Pirani1956}, \cite{Tulczyjew1959} and \cite{Madore1969}. These works differ in the way the multipole moments are defined. In a series of papers \cite{Dixon1970I, Dixon1970II, Dixon1974III}, Dixon and collaborators provide a more thorough definition of the multipole moments. The equations describing the motion of pole--dipole particles are commonly known as the Mathisson--Papapetrou--Dixon (MPD) equations. In general, the analysis leaves the equation of motion of the quadrupole moment unspecified.

In the pole--dipole case, additional equations are needed to define the center of mass, called spin supplementary conditions. Several different spin supplementary conditions have been proposed which lead to different worldlines for the representative point in the body. In the pole--dipole approximation, these worldlines lie within the minimal world tube \cite{Kyrian2007}. Reference \cite{Kyrian2007} gives a list of known spin supplementary conditions and discusses what they imply for conserved quantities for the pole--dipole particle. Reference \cite{Costa2014} gives an in-depth discussion of different spin supplementary conditions.

In this paper, we examine the dynamics of hyperelastic\footnote{A hyperelastic material is an elastic material 
whose stress tensor can be derived from a potential energy function of the strain tensor. For a hyperelastic material, there is no energy dissipation or heat conduction. We sometimes refer to such materials simply as ``elastic."} bodies as models for extended body motion in general relativity. Elastic bodies are closer to physical reality than, for example, rigid bodies. There is difficulty in defining a rigid body in curved spacetime: If a rigid body is defined as one that has no deformation, it would be unphysical because it would require a speed of sound that is greater than the speed of light. Furthermore, stresses generated in the body can be important and contribute to the SEM tensor. In general, it is perhaps simpler to treat extended bodies as elastic (or fluid). If one wants to model stiff bodies, then the material properties of the elastic body can be chosen so that the speed of sound is close to the speed of light.

The motion and deformation of extended bodies in general relativity is an important topic. The quadrupole deformation of neutron stars in binary inspirals can be potentially detected from the observed gravitational waves \cite{Abbott2017,De2018}. These observations should provide crucial insight into the nuclear equation of state of these objects. The deformation, spin, and internal structure of the small body in extreme mass ratio inspirals may have an effect on the gravitational waves emitted \cite{Steinhoff2012}. The planned space interferometer LISA may be able to detect these effects. In particular, the spin of the small body is expected to have a next-to-leading order (i.e., first post-adiabatic order) influence on the phase of these gravitational waves \cite{Witzany2019}. Except for specific cases where the tidal 
field is static \cite{Hinderer_2008} or steady, the deformation of the body cannot be modeled accurately by simply setting it proportional to the tidal field. For example, the small body could be spinning too rapidly to come to equilibrium in response to the tidal forces or might be immersed in the time-changing tidal field in an eccentric or inclined orbit. Thus the treatment of the dynamics of the extended body must expand beyond MPD to include the dynamics of the quadrupole and higher moments, moments which are known \cite{Harte2020} to affect the motion.  

Relativistic hydrodynamics is a very successful theory and is widely used to model fluids in strong gravity and at high Lorentz factors. A major difference between relativistic hydrodynamics and general relativistic elasticity is that shear stresses are absent in perfect fluid hydrodynamics. However, it is known that neutron star crusts are solid \cite{Chamel2008}.  Moreover, some (ultra-massive) white dwarfs are expected to have frozen cores with up to $99\%$ of their mass in crystallized form \cite{Camisassa2019}. As a natural alternative to fluids, elastic bodies allow for shear stresses. 

Numerical works on general relativistic elasticity are few in number. One such work is found in Ref.~\cite{Gundlach2011}. The authors propose an Eulerian formulation of general relativistic elasticity that can be used for numerical modeling and can capture shocks. They test their framework on Riemann problems in Minkowski spacetime. The authors of Refs.~\cite{Alho2022,Alho2023} used general relativistic elasticity to study spherically symmetric elastic stars. They proposed that elasticity might be an important factor for modeling exotic compact objects.
Other works include a set of papers \cite{Karlovini2003I, Karlovini2004II, Karlovini2004III, Karlovini2007IV} that propose a coherent framework for accurately modeling the solid crust within neutron stars. 

In this paper, we are interested in accurately modeling an extended hyperelastic body in general relativity. Our goal is to determine how its motion is affected by its finite size and calculate the changes in its internal structure, including deformation and spin, due to interactions with the background curvature. As a first step, for this work we assume that the extended body's SEM tensor does not affect the spacetime curvature. In other words, we ignore self-gravity and gravitational radiation.  In a paper that will immediately follow, we will show that despite this restriction the system exhibits interesting radiationless self-force effects beyond pole-dipole order, with transfers of energy and angular momentum between an orbit and the body itself. The present paper details the formalism and the numerical method. The elastic body is handled with a Lagrangian scheme, where the mass is broken up into finite elements.  A novel approach to the dynamics is pursued, where the action for the body is spatially discretized. The discrete action in turn leads directly to Euler-Lagrange equations for the finite mass elements as a large set of coupled ordinary differential equations.  The method developed here will be used in future applications that consider extended body encounters with massive black holes, which can be exploited to test MPD and higher-order curvature-coupling effects.

The outline of this paper is as follows. We begin in Sec.~\ref{sec:GRelastic} by reviewing the general relativistic theory of hyperelasticity as formulated in Ref.~\cite{Brown2021}. We explain our numerical method in Sec.~\ref{sec:nummethod}. In Sec.~\ref{sec:testmodels} we rederive and review the normal modes of oscillation for a hyperelastic sphere in the linearized, nonrelativistic limits. We test our code in Sec.~\ref{sec:codeval} by comparing  the numerical and analytical displacements and velocities corresponding to a combination of selected normal modes.

Throughout this paper, we use the sign conventions of Misner, Thorne and Wheeler \cite{MTW}.

\section{General relativistic theory of elasticity}
\label{sec:GRelastic}

In this section, we give a brief review of hyperelasticity theory in general relativity using a Lagrangian formulation as developed in Ref.~\cite{Brown2021}. We focus on the action and the stress--energy--momentum tensor.

The earliest work on generalizing elasticity theory to special relativity is by Herglotz \cite{Herglotz1911}. Subsequently, DeWitt \cite{DeWitt:1962cg} extended Herglotz' theory to the general relativistic domain to describe a ``stiff elastic medium''. He used this structure to aid in the formulation of a quantum theory of gravity. Later works on general relativistic elasticity theory include Carter and Quintana \cite{Carter1972}, Kijowski and Magli \cite{Kijowski1992}, Beig and Schmidt \cite{Beig2003, Beig_2017}, Gundlach, Hawke and Erickson \cite{Gundlach2011} and Beig \cite{Beig2023}.

Some of these works favor an Eulerian formulation while the others use a Lagrangian approach. In the Eulerian approach, the fundamental variables are fields on spacetime. In the Lagrangian approach, the fundamental variables are time--dependent fields on ``matter space", the space of material particles that make up the elastic body. The two approaches are mathematically equivalent. The advantage of the Lagrangian formulation for numerical modeling is that it is easier to implement natural boundary conditions \cite{Lanczos1949} where the surface is free to move. In the Eulerian formulation the surface is not simply defined and requires interpolation. Also, since the Lagrangian field equations are formulated on matter space rather than physical space, the number of (discrete) equations to be solved is much smaller in the Lagrangian approach.

\subsection{World tube, radar metric and Lagrangian strain}

Let the four--dimensional spacetime manifold be denoted by $\mathcal{M}$, with spacetime coordinates $x^\mu$ and metric $g_{\mu\nu}$. The matter space, $\mathcal{S}$, is the space of material points with coordinates $\zeta^i$ for $i=1,2,3$. (Note that Latin indices beginning with $i,j,k,\ldots$ should not be confused with the indices of the spatial subset of spacetime coordinates.)

\begin{figure}
\centering
\includegraphics{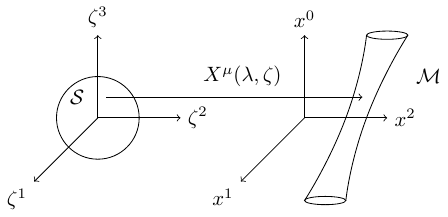}
\caption{\label{fig:Mapping} $X^\mu(\lambda,\zeta)$ maps a material point with coordinates $\zeta^i$ in $\mathcal{S}$ and a real parameter $\lambda$ to the spacetime event $x^\mu$ inside the world tube in four--dimensional spacetime, $\mathcal{M}$.}
\end{figure}

Let $\lambda$ be a real parameter. The functions $X^\mu(\lambda, \zeta)$ are maps from $\mathfrak{R} \times \mathcal{S}$ to $\mathcal{M}$, see Fig.~\ref{fig:Mapping}. As $\lambda$ is continuously varied, $X^\mu(\lambda, \zeta)$ traces the timelike worldline of the material point $\zeta^i$. The collection of all worldlines corresponding to the material points of the body is called the world tube.

The four--velocity of a material point is
\be 
U^\mu = \dot X^\mu/\alpha \ , 
\ee 
where the ``dot" denotes $\partial/\partial\lambda$ and
\be 
\alpha = \sqrt{-\dot X^\mu \dot X_\mu}\ ,
\ee  
is the material lapse function. 
The radar metric, $f_{\mu\nu}$, defined inside the world tube is
\be 
f_{\mu \nu} = g_{\mu\nu} + U_\mu U_\nu\ .
\ee 
The name ``radar'' comes from Landau and Lifshitz \cite{landau-lifshitz-course} who used light signals to find the spatial distance between two infinitesimally separated events. It is easy to see that $f_{\mu\nu} U^\mu = 0$ and that $f^\nu_\mu V^\mu$ is orthogonal to $U^\mu$ for any vector $V^\nu$. Hence, $f_\mu^\nu$ is a ``projection tensor" that projects $V^\mu$ into the space orthogonal to $U^\mu$. The radar metric can be mapped back to the matter space,
\be 
f_{ij} = X^\mu_{,i} f_{\mu\nu} X^\nu_{,j}\ .
\ee 
where $,j$ denotes $\partial/\partial \zeta^j$. The radar metric $f_{ij}$ gives distances between infinitesimally separated material points such that the distance is measured in the rest frame of the points in physical spacetime, $\mathcal{M}$. That is, $ds^2 = f_{ij} d\zeta^i d\zeta^j$, is the square of the proper distance between material points. 

The Lagrangian strain tensor can be defined in the same way as in nonrelativistic elasticity:
\be 
E_{ij} = (f_{ij} - \epsilon_{ij})/2\ . 
\ee 
Here, $\epsilon_{ij}$ is the ``relaxed metric" on matter space. That is, $\epsilon_{ij} d\zeta^i d\zeta^j$ is the square of the physical distance between nearby material points when the body is undeformed.

The deformation gradient gives the amount of strain in the material and in the relativistic domain it is defined using the radar metric and the map, $X^\nu_{,i}$,
\be 
F_{\mu i}= f_{\mu\nu} X^\nu_{,i}\ .
\ee
Another important tensor is the second Piola--Kirchhoff stress tensor\footnote{Various measures of stress in the nonrelativistic domain are 
described in works on continuum mechanics, such as Bower \cite{Bower} and Kelly \cite{Kelly}. The first and 
second Piola--Kirchhoff stress tensors were introduced by Piola \cite{Piola} and Kirchhoff \cite{Kirchhoff}.} defined as the gradient of the energy density $\rho$ with respect to the Lagrangian strain,
\be\label{stresstensordefined}
S^{ij} = \frac{\pd \rho}{\pd E_{ij}}\ .
\ee

\subsection{Action and stress--energy--momentum tensor}

Hyperelastic materials have a stored energy function that can be specified in terms of the Lagrangian strain $E_{ij}$. The energy density per unit of undeformed volume is denoted by $\rho(E)$ and is a function of $E_{ij}$ and $\epsilon_{ij}$. It can also depend on $\zeta^i$ if the material is not uniform. The dependence on $\epsilon_{ij}$ and $\zeta$ has been omitted in the notation to make it more compact. The relativistic action for a hyperelastic body is \cite{DeWitt:1962cg, Brown1996, Brown2021}, 
\be 
S[X,g] = - \int_{\lambda_i}^{\lambda_f} d\lambda \int_{\mathcal{S}}d^3\zeta \sqrt{\epsilon} \, \alpha \rho\ . \label{eq:action}
\ee 
This action is a generalization of the action for a continuum of particles with ``nearest neighbor" interactions mediated by the Lagrangian strain tensor.

The energy density can be written as
\be 
\rho(E)= \rho_0 + W(E)\ ,
\ee 
where $\rho_0$ is the rest mass per unit undeformed volume and $W$ is the potential energy (or interaction energy) per unit undeformed volume.
The interaction energy of the hyperelastic body is obtained by using distances computed in the rest frames of elements of the body. 

Let $x^0 \equiv t = \mathrm{const}$ correspond to spacelike hypersurfaces and let $x^a$ denote the spatial  subset of the spacetime coordinates, where $a=1,2,3$. The coordinate basis vectors $\pd/\pd x^a$ are spacelike. Because of the gauge invariance of the action, we can freely choose the parameter $\lambda$ along each worldline. Thus, we can choose the parameterization $\lambda = x^0 \equiv t$. Then, $\dot X^0 = 1$ and $X^0_{,i}=0$. With this gauge choice the action in the $\lambda = t$ gauge can be written as
\be
    S[X] = -\int_{t'}^{t''} dt \,  \int_{\mathcal S} d^3\zeta \, 
    \sqrt{\epsilon}\alpha\rho(E) \ , \label{eq:actiontgauge}
\ee
which is a functional of $X^a(t,\zeta)$. 

In this gauge the radar metric and material lapse are
\bse
\begin{align}
    f_{ij} &= X^a{}_{,i}(g_{ab} + \gamma^2 V_a V_b) X^b{}_{,j} \\
    \alpha &= N \sqrt{1 - V^a V_a}
\end{align}
\ese
where the ``dot" now denotes $\partial/\partial t$. Here,  $g_{ab}$ is the spatial metric and 
\be
    V^a \equiv (\dot X^a + N^a)/N
\ee
with $N = \sqrt{-1/g^{tt}}$ denoting 
the spacetime lapse function and $N_a = g_{ta}$ denoting the shift vector. 
The spatial vector $V^a$ is the velocity of the 
material as seen by observers at rest in the
$t = {\rm const}$ surfaces. We have also defined the Lorentz 
factor $\gamma \equiv 1/\sqrt{1 - V^a V_a}$. 
Note that spatial indices are raised and lowered with the spatial metric.

The stress--energy--momentum (SEM) tensor for matter fields is obtained from the functional derivative of the matter action with respect to the metric, 
\be
T^{\mu\nu} (x) = \frac{2}{\sqrt{-g}}\frac{\delta S_{\mathrm{matter}}}{\delta g_{\mu\nu}(x)}\ .
\ee 
The final form of the SEM tensor is \cite{Brown2021}
\be
T^{\mu\nu}(X(\lambda,\zeta)) =  \frac{1}{J}\, [\rho U^\mu U^\nu +  S^{ij}F^\mu_i F^\nu_j]\ , \label{eq:SEM}
\ee 
where $J\equiv\sqrt{f}/\sqrt{\epsilon}$. The metric $\epsilon_{ij}$ gives distances between material points $\zeta^i$ in $\mathcal{S}$ when the elastic body is relaxed and $f_{ij}$ gives distances between material points $\zeta^i$ in $\mathcal{S}$ when the elastic body is deformed. Therefore, the factor $1/J$ converts energy density per unit undeformed volume to per unit deformed volume. The SEM tensor satisfies local conservation, $\nabla_\mu T^{\mu\nu}=0$.

\section{Numerical method}
\label{sec:nummethod}

Numerical methods for solving partial differential equations (PDEs) include finite difference (FD), finite volume (FV) and finite element (FE) methods. FE methods are particularly useful in solving elasticity problems. By using triangular or tetrahedral meshes, they allow boundaries of elastic bodies to be represented more closely than the rectangular grids used in FD and FV methods. In FD methods, the PDEs are discretized directly whereas in FV methods, the PDEs are integrated over a volume element. In FE methods, the PDEs are converted to a weak form by multiplying with a test function that satisfies the boundary conditions and then integrating over the domain \cite{Li2017}. 

We discretize the action of the elastic body directly instead of discretizing the partial differential equations of motion. This leads to the free surface or natural boundary condition where variations at the boundary are nonzero, to be trivially implemented via the variational process. We use FE methods with tetrahedral elements to model elastic bodies of any shape such as spheres or ellipsoids. These models can be used to describe solid astrophysical objects. We discretize the action in space and not in time and obtain ordinary differential equations (ODEs) in mass matrix form. There is a suite of well-tested methods that can be used to solve such coupled ODEs. 

We use Matlab's partial differential equation toolbox \cite{pdetool} to generate a linear tetrahedral mesh for three-dimensional bodies. To utilize computing clusters, we use the software package Metis \cite{Metis} to partition the mesh and parallelize the algorithm. The Message Passing Interface (MPI) is used to communicate neighbor information.

\subsection{Matter space discretization}

\begin{figure}
\centering
\includegraphics{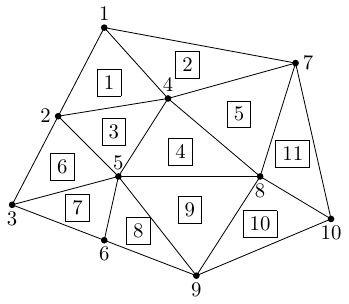}
\caption{\label{fig:elementsandnodes}Here we are depicting a two-dimensional triangular mesh instead of a tetrahedral mesh for clarity. This figure shows node labels, $n=\{1,2..,10\}$ and element labels $E=\{1,2..,11\}$ (boxed). The set of nodes in element, $E=4$, is $\mathcal{N}(4) = \{4,5,8\}$. The ring of node $n=5$ is ${\mathcal R}(5)=\{3,4,6,7,8,9\}$.} 
\end{figure}

The matter space $\mathcal S$ is divided into non-overlapping elements. Let ${\mathcal S}_E$ for $E = 1,2,\ldots$ denote the elements, that is, ${\mathcal S}$ is the union of the ${\mathcal S}_E$'s. Let $n = 1,2,\ldots$ label the nodes throughout the body. Each node in the body has a unique index number. Let ${\mathcal N}(E)$ denote the set of nodes in element $E$. An example of ${\mathcal N}(E)$ is shown in Fig.~\ref{fig:elementsandnodes}. Then, for $\zeta^i \in {\mathcal S}_E$, we have
\be
    X^a(t,\zeta) = \sum_{n\in \mathcal{N}(E)} X^a_{n}(t) \; \phi^E_n(\zeta)\ , \quad \zeta^i \in \mathcal{S}_E\ , \label{eq:discX}
\ee 
where the sum is over the nodes contained in the element $\mathcal{S}_E$. Note that the shape functions $\phi^E_n(\zeta)$ depend on the node as well as the element. 

\subsection{Semi-discretized action and equations of motion}

The action in the $\lambda=t$ gauge (Eq.~(\ref{eq:actiontgauge})) is discretized using Eq.~(\ref{eq:discX}),
\begin{align} 
    S[X] &= \int_{t'}^{t''} dt \sum_E  \int_{\mathcal{S}_E} d^3\zeta \; \mathcal{L} \bigg(\sum_{n\in \mathcal{N}(E)} X^a_{n}(t) \; \phi_n^E(\zeta), \nono \\
    &\sum_{n\in \mathcal{N}(E)} \dot X^a_{n}(t) \; \phi_n^E(\zeta), \; \sum_{n\in \mathcal{N}(E)} X^a_{n}(t) \; \phi^E_{n,i} \bigg)\ , \label{eq:discaction}
\end{align}
where the Lagrangian density is defined by 
${\mathcal L}(X,\dot X,X_{,i}) = -\sqrt{\epsilon}\alpha\rho(E)$. The action is a functional of the coordinates of each node,  $X^a_{n}(t)$.

\begin{figure}
\centering
\includegraphics{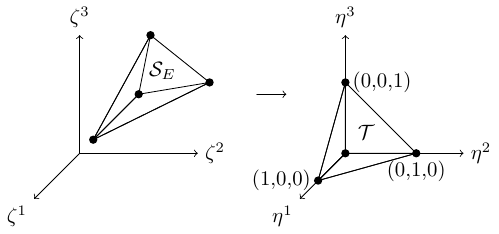}
\caption{\label{fig:transformation} A general tetrahedral element $\mathcal{S}_E$ in $\mathcal{S}$ is transformed into a unit trirectangular tetrahedron ${\mathcal{T}}$ with a node at origin and the three other nodes displaced by one unit along the coordinate axes. It does not matter which node is at the origin.} 
\end{figure}

We select the element type to be linear tetrahedrons with nodes at the vertices only. A general tetrahedral element $\mathcal{S}_E$ is transformed into a unit trirectangular tetrahedron ${\mathcal{T}}$ with coordinates $\eta^i$. Let $\zeta^i_{(\alpha)}$ denote the coordinates of the four nodes, for $\alpha = 0,1,2,3$. 
The transformation is linear, with $\zeta^i = A^{ij} \eta^j + B^i$ where $A^{ij}$ and $B^i$ are constants in each element. These constants are given by
\bse
\begin{align}
    B^i &= \zeta^i_{(0)} \ ,\\
    A^{i1} &= \zeta^i_{(1)} - \zeta^i_{(0)}  \ ,\\
    A^{i2} &= \zeta^i_{(2)} - \zeta^i_{(0)}  \ ,\\
    A^{i3} &= \zeta^i_{(3)} - \zeta^i_{(0)}  \ .
\end{align}
\ese
In the new coordinates, the nodes have 
coordinates $\eta^i_{(0)} = (0,0,0)$, $\eta^i_{(1)} = (1,0,0)$, $\eta^i_{(2)} = (0,1,0)$, and $\eta^i_{(3)} = (0,0,1)$. Figure \ref{fig:transformation} shows the transformation.

Let $\alpha(n)$ map the four node numbers of ${\mathcal S}_E$ to 
the set $\{0,1,2,3\}$. The shape function defined in terms of the new coordinates $\eta^i$ are
\be
    \bar\phi_{\alpha(n)}(\eta) \equiv \phi^E_n(\zeta(\eta))\ .
\ee
Explicitly, the linear shape functions are given by (see Eqs.~(3.1.19)--(3.1.22) of Ref.~\cite{hughes2012finite})
\bse
\begin{align}
\bar\phi_{0}(\eta ) &= 1-\eta^1-\eta^2-\eta^3\ , \\
\bar\phi_{1}(\eta) &= \eta^1\ , \\
\bar\phi_{2}(\eta) &= \eta^2\ , \\
\bar\phi_{3}(\eta) &= \eta^3 \ .   
\end{align}
\ese
In the new coordinates, the action is
\begin{align}
    S[X] &= \int_{t'}^{t''} dt \sum_E
    \int_{{\mathcal T}} d^3\eta \, |J_{E}|
    \mathcal{L}\bigg(\sum_{n\in {\mathcal N}(E)} X^a_{n}(t) \; \bar\phi_{\alpha(n)}(\eta), \nono \\
    &\sum_{n\in{\mathcal N}(E)} \dot X^a_{n}(t) \; \bar\phi_{\alpha(n)}(\eta), \sum_{n\in \mathcal{N}(E)} X^a_{n}(t) \; \phi^E_{n,i}  \bigg)\ ,
\end{align}
where $|J_{E}|$ is the determinant of the Jacobian of the transformation from $\zeta^i$ to $\eta^i$ for element $E$. 
We can pull $|J_{E}| $ outside the integral since it is independent of $\eta^i$. It should be noted that $\phi^E_{n,i} \equiv 
\partial \phi^E_n/\partial \zeta^i$ are constants, independent of $\eta^i$.

We now replace the integral over $\eta^i$ in each element with a quadrature rule:
\begin{align}
    S[X] &= \int_{t'}^{t''} dt \sum_E 
    \sum_\sigma w_\sigma  \, |J_{E}|
    \mathcal{L}\bigg(\sum_{n\in {\mathcal N}(E)} X^a_{n}(t) \; \bar\phi_{\alpha(n)}(\eta_{(\sigma)}), \nono \\
    &\qq\sum_{n\in{\mathcal N}(E)} \dot X^a_{n}(t) \; \bar\phi_{\alpha(n)}(\eta_{(\sigma)}), \sum_{n\in \mathcal{N}(E)} X^a_{n}(t) \; \phi^E_{n,i}  \bigg)\ ,
\end{align}
for some set of points $\eta^i_{(\sigma)}$ in ${\mathcal T}$. We choose the points to coincide with the nodes (vertices) of the element, and choose weights $w_\sigma = 1/24$ for each node. With this weighting, the integration of linear functions is exact. 

Using the results $\bar\phi_\alpha(\eta_{(\sigma)}) = \delta_{\alpha\sigma}$, the discrete action becomes
\begin{align}
    S[X] &= \int_{t'}^{t''} dt \sum_E 
    \sum_\sigma w_\sigma  \, |J_{E}|
    \mathcal{L}\bigg(\sum_{n\in {\mathcal N}(E)} X^a_{n}(t) \;\delta_{\alpha(n)\sigma}, \nono \\
    &\sum_{n\in{\mathcal N}(E)} \dot X^a_{n}(t) \; \delta_{\alpha(n)\sigma}, \; \sum_{n\in \mathcal{N}(E)} X^a_{n}(t) \; \phi^E_{n,i}  \bigg)\ .
\end{align}
For each value of $\sigma$ in the sum, the only term in the 
first argument of ${\mathcal L}$ that is nonzero is the one 
for which $\alpha(n) = \sigma$. Likewise for the second argument of ${\mathcal L}$. Thus, we can write the action as 
\begin{align}
    S[X] &= \frac{1}{24} \int_{t'}^{t''} dt \sum_E 
    \sum_{n\in \mathcal{N}(E)} |J_{E}| {\mathcal L} \bigg( X^a_{n}(t), \; \dot X^a_{n}(t), \nono\\
    &\qq\qq\qq\qq\sum_{m\in \mathcal{N}(E)} X^a_{m}(t) \; \phi^E_{m,i} \bigg)\ .
\end{align} 

Let ${\mathcal R}(n)$ be the ``ring" of $n$. This is the list of elements ($E$ values) that have $n$ as one of their nodes. An example of ${\mathcal R}(n)$ is shown in Fig.~\ref{fig:elementsandnodes}. We isolate the terms in the action that involve the variable $X^a_N$ for some fixed node number $N$. Let these terms be denoted by $S_N$: 
\begin{align}
    S_N &= \frac{1}{24} \int_{t'}^{t''} dt \sum_{E\in {\mathcal R}(N)} 
    \sum_{n\in \mathcal{N}(E)} |J_{E}| {\mathcal L} \biggl( X^a_{n}(t), \; \dot X^a_{n}(t), \nono\\
    &\qq\qq\qq\qq\q\sum_{m\in \mathcal{N}(E)} X^a_{m}(t) \; \phi^E_{m,i}  \biggr)\ .
\end{align}
Only elements in the ring of $N$ depend on the node $X^a_N$. In the sum over nodes for each element, there are two cases. One case is when the node number $n$ equals $N$, the other is when $n$ does not equal $N$. Therefore, we find
\begin{widetext}
\begin{align}
S_N & =  \frac{1}{24} \int_{t'}^{t''} dt \sum_{E\in {\mathcal R}(N)}  |J_{E}|\biggl\{ 
    {\mathcal L} \biggl( X^a_{N}(t), \; \dot X^a_{N}(t), \sum_{m\in \mathcal{N}(E)} X^a_{m}(t) \; \phi^E_{m,i} \biggr) \nono \\
    &\qq\qq\qq\qq\qq\qq\qq\qq\qq\qq\q+ \sum_{n\in \mathcal{N}(E), n\ne N} 
     {\mathcal L} \biggl( X^a_{n}(t), \; \dot X^a_{n}(t), \sum_{m\in \mathcal{N}(E)} X^a_{m}(t) \; \phi^E_{m,i} \biggr) \biggr\}\ .   
\end{align}
It should be noted that $X^a_N$ occurs in the third argument of ${\mathcal L}$ in both terms. 

We now vary $S_N$ with respect to $X^a_N$: 
\begin{align}
\delta S_N & = \nono\\
&\frac{1}{24} \int_{t'}^{t''} dt \sum_{E\in {\mathcal R}(N)} |J_{E}| \biggl\{ 
    \frac{\partial{\mathcal L}}{\partial X^a}\biggr|_{N,E} \delta X^a_N + \frac{\partial{\mathcal L}}{\partial \dot X^a}\biggr|_{N,E} \delta \dot X^a_N 
    + \frac{\partial{\mathcal L}}{\partial  X^a_{,i}}\biggr|_{N,E} \phi^E_{N,i} \delta X^a_N + \sum_{n\in \mathcal{N}(E), n\ne N} \frac{\partial{\mathcal L}}{\partial X^a_{,i}}\biggr|_{n,E} \phi^E_{N,i} \delta X^a_N \biggr\}\ ,    
\end{align}
where the symbol $|_{n,E}$ indicates that the partial derivatives are evaluated at $X^a = X^a_{n}$, $\dot X^a =  \dot X^a_{n}$, and $X^a_{,i} = \sum_{m\in \mathcal{N}(E)} X^a_{m}(t) \phi^E_{m,i}$. The last two terms in $\delta S_N$ can be combined into a single sum over all $n\in \mathcal{N}(E)$. Then the functional derivative (Lagrange's equation) is 
\begin{align}
    0 &= \frac{\delta S}{\delta X^a_N} = \frac{1}{24}
    \sum_{E\in {\mathcal R}(N)} |J_{E}| \biggl\{ 
    \frac{\partial{\mathcal L}}{\partial X^a}\biggr|_{N,E}  - \frac{d}{dt} \left( \frac{\partial{\mathcal L}}{\partial \dot X^a}\biggr|_{N,E} \right) + \sum_{n\in \mathcal{N}(E)} \frac{\partial{\mathcal L}}{\partial  X^a_{,i}}\biggr|_{n,E} \phi^E_{N,i} 
    \biggr\}\ .    
\end{align}
Next, we expand the total time derivative: 
\begin{align}
    \frac{d}{dt}\left( \frac{\partial{\mathcal L}}{\partial \dot X^a}\biggr|_{N,E}\right)
    &= \frac{\partial^2{\mathcal L}}{\partial \dot X^b \partial \dot X^a} \biggr|_{N,E} \ddot X^b_N 
    + \frac{\partial^2{\mathcal L}}{\partial X^b \partial\dot X^a}\biggr|_{N,E} \dot X^b_N + \frac{\partial^2{\mathcal L}}{\partial X^b_{,i} \partial\dot X^a} \biggr|_{N,E} \sum_{m\in \mathcal{N}(E)} \dot X^b_m \phi^E_{m,i}\ .  
\end{align}
Then the equations of motion are 
\begin{align} \label{bigMequalsFequation}
    \underbrace{\sum_{E\in {\mathcal R}(n)} |J_{E}| \frac{\partial^2{\mathcal L}}{\partial \dot X^b \partial \dot X^a} \biggr|_{n,E}}_{(M_{ab})_n} \ddot X^b_n =  \underbrace{\begin{aligned}[t]
    &\sum_{E\in {\mathcal R}(n)} |J_{E}| \bigg\{ 
    \frac{\partial{\mathcal L}}{\partial X^a}\biggr|_{n,E}  + \sum_{m\in \mathcal{N}(E)} \frac{\partial{\mathcal L}}{\partial  X^a_{,i}}\biggr|_{m,E} \phi^E_{n,i}     
    - \, \frac{\partial^2{\mathcal L}}{\partial X^b \partial\dot X^a}\biggr|_{n,E} \dot X^b_n \\
    &\qq\qq\qq\qq\qq\qq\qq\qq\qq- \, \frac{\partial^2{\mathcal L}}{\partial X^b_{,i} \partial\dot X^a} \biggr|_{n,E} \sum_{m\in \mathcal{N}(E)} \dot X^b_m \phi^E_{m,i}  
    \bigg\}\ , \end{aligned}}_{(F_a)_n} 
\end{align}
where $n$ is replaced by $m$ and $N$ is 
replaced by $n$.

For each value of $n$ in Eq.~(\ref{bigMequalsFequation}), the coefficient of $\ddot X^b_n$ is a $3\times 3$ matrix in the indices $a$ and $b$.  These equations are rewritten as a system of $6N_{\mathrm{total}}$ first order ODEs for the variables $X^a_n$ and $V^a_n = \dot X^a_n$, where $N_{\mathrm{total}}$ is the 
total number of nodes. The first $3N_{\mathrm{total}}$ equations are the definitions  $V^a_n = \dot X^a_n$ with $a=1,2,3$ and $n=1,\ldots,N_{\mathrm{total}}$. Denoting the coefficient of $\Ddot{X}^b_n$ in Eq.~(\ref{bigMequalsFequation}) as $(M_{ab})_n$ and the right hand side as $(F_a)_n$, the next $3N_{\mathrm{total}}$ first order ODEs are written in matrix form as
\begin{align}    
\underbrace{\begin{bmatrix}
(M_{11})_1 & (M_{12})_1 & (M_{13})_1 & \ldots&0\\
(M_{21})_1 & (M_{22})_1 & (M_{23})_1 & \ldots&0\\
(M_{31})_1 & (M_{32})_1 & (M_{33})_1 & \ldots&0\\
\vdots\\
0          &   \ldots   &   \ldots   &  \ldots
\end{bmatrix}}_{\text{mass matrix},\, M}
\frac{d}{d t} \left(
\begin{bmatrix}
V^1_{1}\\
V^2_{1}\\
V^3_{1}\\
\vdots\\
\vdots
\end{bmatrix}\right)
=
\underbrace{\begin{bmatrix}
(F_1)_1\\
(F_2)_1\\
(F_3)_1\\
\vdots\\
\vdots\\
\end{bmatrix}}_{\text{vector},\, F}\ . \label{eq:odemat}
\end{align}
\end{widetext}
The mass matrix $M$ is pentadiagonal. We use the subroutine DGBSV from the Fortran Linear Algebra Package (LAPACK) which uses lower–upper (LU) decomposition to solve the linear system of equations (\ref{eq:odemat}) and obtain the time derivatives of $V^a_n$. We then use the fourth--order Runge--Kutta scheme to evolve $X^a_n$ and $V^a_n$ at discrete values of $t$.

From numerical experiments we find that the Courant condition,
\be 
\Delta t \leq h_\mathrm{min}/C_L\ ,
\ee
must be met for stability. Here, $\Delta t$ is the time step size, $h_\textrm{min}$ is the minimum edge length of the tetrahedral elements, and $C_L$ is the longitudinal sound speed (see Eq.~(\ref{eq:CLCT}) defined below) which is the maximum sound speed in the material.

\section{Test models in the nonrelativistic domain}
\label{sec:testmodels}

Elasticity theory in the nonrelativistic domain has a long history and many applications. A large body of works make use of linear elasticity for which exact solutions are known in some cases. Here we reproduce the exact solutions for the normal mode oscillations of a free solid elastic sphere. In Sec.~\ref{sec:codeval} we use our relativistic, nonlinear code to simulate the motion of a solid elastic sphere in flat spacetime, and show that the expected results are obtained in the limit of small, nonrelativistic oscillations. 

Nonrelativistic elasticity theory can be obtained from general relativistic elasticity theory (Sec.~\ref{sec:GRelastic}) by taking the nonrelativistic limit, as shown in Ref.~\cite{Brown2021}. The resulting action, deduced from (\ref{eq:actiontgauge}), is
\be\label{NonRelAction}
S[X] = \int_{t_i}^{t_f}dt\int d^3\zeta \,\sqrt{\epsilon}\,\bigg[\frac{1}{2}\rho_0 \dot X^a \dot X_a - W(E) - \rho_0 \Phi\bigg]\ ,
\ee 
where $\Phi$ is the Newtonian gravitational potential. The index on $\dot X_a$ has been lowered with the spatial metric $g_{ab}$ which in this section is taken to be flat. Here, the energy density 
is written as 
\be 
\rho(E) = \rho_0 + W(E)\ ,
\ee 
where $\rho_0$ is the rest mass density per unit undeformed volume and $W(E)$ is the potential energy density per unit undeformed volume. 
In this nonrelativistic limit the radar metric reduces to
\be 
f_{ij} = X^a_{,i}g_{ab}X^b_{,j}\ .
\ee
The second Piola stress tensor, $S^{ij}$, is  the derivative of $W(E)$ with respect to the Lagrangian strain, $E_{ij}=(f_{ij} - \epsilon_{ij})/2$, as in Eq.~(\ref{stresstensordefined}). 

\subsection{Hyperelastic energy models}
\label{sec:elasener}
Elastic materials are materials for which the stress can be written in terms of the strain at a particular time. Hyperelastic materials are materials for which the work done by stresses during the deformation process depends only on the initial and final configurations. Homogeneous materials are materials for which portions of the elastic material have the same mechanical behaviour. Isotropic materials are materials for which the potential energy function, $W$, depends on the deformation gradient only through $f_{ij}$ and $\epsilon_{ij}$. Some energy models \cite{Brown2021} for isotropic hyperelastic materials include the Saint Venant-Kirchhoff, the Mooney-Rivlin \cite{Mooney2004}, the neo-Hookean and the Ogden models \cite{Ogden1972}.

In this paper we use the Saint Venant-Kirchhoff model with potential energy function
\be 
W(E) = \frac{\lambda}{2}(\epsilon^{ij} E_{ij})^2 + \mu(\epsilon^{ik}\epsilon^{jl}E_{ij}E_{kl})\ .
\ee 
Here, $\epsilon^{ij}$ is the inverse of $\epsilon_{ij}$ and $\lambda$ and $\mu$ are the Lam\'e constants. (The Lam\'e constant $\lambda$ should not to be confused with our previous use of $\lambda$ as a path parameter in Sec.~\ref{sec:GRelastic}.) The bulk modulus, $K= \lambda + 2\mu/3$, measures resistance to volume changes. The Saint Venant-Kirchhoff model is not valid for large strains because the model softens under large compression.

\subsection{Linear elasticity}
\label{subsec:linelas}
Linear elasticity is used when the deformation is the result of small displacements from some reference configuration, which we denote by $X^a_R(\zeta)$. We also assume that there is no rotation. We then write
\be 
X^a(\zeta, t) = X^a_R(\zeta) + \xi^a(\zeta, t)\ ,
\ee 
where $|\xi^a(\zeta, t)|$ is small. 
Then we have
\begin{align}
\dot X^a(\zeta, t) &= \dot \xi^a(\zeta, t)\ , \\
X^a_{,i}(\zeta, t) &= X^a_{R,i}(\zeta) + \xi^a_{,i}(\zeta, t) \ ,
\end{align}
and we also assume that $|\xi^a_{,i}(\zeta, t)|$ is small. Choosing flat space and Cartesian coordinates, the radar metric (also known as the right Cauchy-Green deformation tensor) and the relaxed matter space metric become
\begin{align}
f_{ij} &= X^a_{,i} \delta_{ab} X^b_{,j} = (X^a_{R,i}+\xi^a_{,i})\delta_{ab}(X^b_{R,j}+\xi^b_{,j})\ , \label{eq:fijlinear}\\
\epsilon_{ij} &= X^a_{R,i} \delta_{ab} X^b_{R,j}\ . \label{eq:epsijlinear}
\end{align}
From the map $x^a = X^a_R(\zeta)$ we can define the inverse map that takes a point in physical space to the matter space label for the body in its relaxed state:
\be
\zeta^i = Z^i_R(x)\ .
\ee 
Differentiation with respect to $\zeta$ yields the useful relations: 
\begin{align}\label{XZidentities}
X^a_{R,i}Z^j_{R,a} &= \delta^j_i\ , \\
X^a_{R,i}Z^i_{R,b} &= \delta^a_b\ .   
\end{align}
The following formulas for the matter space metric and its inverse hold:
\begin{align}\label{epsionidentities}
\epsilon^{ij} &= Z^i_{R,a} \delta^{ab} Z^j_{R,b}\ ,\\
\delta^{ab} &= X^a_{R,i} \epsilon^{ij} X^b_{R,j}\ , \\
\delta_{ab} &= Z^i_{R,a} \epsilon_{ij} Z^j_{R,b}\ .
\end{align}
We can verify these by computing  $\epsilon^{ij}\epsilon_{jk} = \delta^i_k$  and $\delta^{ab}\delta_{bc}=\delta^a_c$.

We now linearize the Saint Venant-Kirchhoff energy model by expanding $W(E)$ to second order in $\xi^a_{,i}$. Insert Eq.~\ref{eq:fijlinear} and Eq.~\ref{eq:epsijlinear} into the Lagrangian strain tensor $E_{ij} = (f_{ij}-\epsilon_{ij})/2$ to obtain
\be
E_{ij} 
= \frac{1}{2} [\xi^a_{,i}\delta_{ab}X^b_{R,j} + X^a_{R,i}\delta_{ab}\xi^b_{,j}] +{\cal O}^2(\xi^a_{,i})\ .    
\ee
Using the identities (\ref{XZidentities})--(\ref{epsionidentities}) above, we find 
\be\label{TraceepsilonE1}
\epsilon^{ij}E_{ij} =Z^i_{R,a}\xi^a_{,i}+{\cal O}^2(\xi^a_{,i})\ .
\ee
With a slight abuse of notation, we can define $\xi^a(x) \equiv \xi^a(Z_R(x))$ so that $Z^i_{R,b} \xi^a_{,i} = \xi^a_{,b}$. Then the 
result (\ref{TraceepsilonE1}) becomes
\be\label{TraceepsilonE}
\epsilon^{ij}E_{ij} = \xi^a_{,a} +{\cal O}^2(\xi^a_{,i}) \ .
\ee
A similar calculation gives 
\be
\epsilon^{ik}\epsilon^{jl}E_{ij}E_{kl} =\frac{1}{2}[\xi^d_{,e}\xi^e_{,d}+\xi^d_{,e}\xi_{d,}{}^e]+{\cal O}^4(\xi^a_{,i})\ .
\ee
Then to second order in  $\xi^a$ and its derivatives, we obtain 
\be 
W(E)=\frac{\lambda}{2}(\xi^a_{,a})^2 + \frac{\mu}{2}(\xi^d_{,e}\xi^e_{,d}+\xi^d_{,e}\xi_{d,}{}^e)\ , 
\ee 
for the Saint Venant-Kirchhoff model.

\subsection{Dynamical solution for normal modes of an elastic sphere}
\label{subsec:dynsol}

Equation (\ref{NonRelAction}) gives the action for an elastic body in nonlinear elasticity. We specialize to free oscillations by setting the gravitational 
potential to zero, $\Phi = 0$. We specialize to the linear Saint Venant-Kirchhoff model by using the results of the previous subsection. These results assume a flat spatial metric with Cartesian coordinates, so that $g_{ab} = \delta_{ab}$. Then the action becomes
\begin{align}
S[\xi] &= \int_{t_i}^{t_f} dt \int_{\mathcal{S}} d^3\zeta \,\sqrt{\epsilon}\,\bigg[\frac{1}{2} \rho_0 \dot \xi^a \delta_{ab}\dot \xi^b - \frac{\lambda}{2}(\xi^a_{,a})^2 \nono\\
&\qq\qq\qq\qq\q- \frac{\mu}{2}(\xi^d_{,e}\xi^e_{,d}+\xi^d_{,e}\xi_{d,}{}^e)\bigg]\ .
\end{align} 
We can transform the matter space integral over $d^3 \zeta$ to a physical space integral over $d^3x$ using the Jacobian of the transformation $\vert \mathrm{det}(X^a_{R,i})\vert = 1/\sqrt{\epsilon}$. Thus, we find
\begin{align}
S[\xi] &= \int_{t_i}^{t_f} dt \int_{\mathcal{R}} d^3x \bigg[\frac{1}{2} \rho_0 \dot \xi^a \delta_{ab}\dot \xi^b - \frac{\lambda}{2}(\xi^a_{,a})^2 \nono\\
&\qq\qq\qq\qq\q- \frac{\mu}{2}(\xi^d_{,e}\xi^e_{,d}+\xi^d_{,e}\xi_{d,}{}^e)\bigg]\ ,
\end{align} 
where $\mathcal{R}$ is the spatial extent of the undeformed body.

The variation of the action is
\begin{align}
\delta S &= \int_{t_i}^{t_f} dt \int_{\mathcal{R}} d^3x \bigg[-\rho_0 \Ddot \xi^a \delta_{ac} + \lambda \xi^a_{,a,d}\delta^d_c \nono\\
&\qq\qq\qq\qq\qq+ \mu (\xi^d_{,c,d}+\xi_{c,}{}^d{}_{,d}) \bigg] \delta\xi^c \notag \\
 &- \int_{t_i}^{t_f} dt \int_{\partial\mathcal{R}} d^3x \bigg[\lambda \xi^a_{,a} \delta^d_c + \mu(\xi^d_{,c}+\xi_{c,}{}^d)\bigg] \delta\xi^c n_d\ ,
\end{align}
where $n_c$ is the normal to the boundary. In deriving this result, we have integrated by parts to remove derivatives on $\delta \xi^a$ and used the fact that variations vanish at the initial and 
final times, $t_i$ and $t_f$. Setting $\delta S=0$, we find the bulk equations 
\be
-\rho_0 \Ddot \xi_c  + \lambda \xi^a_{,a,c} + \mu (\xi^d_{,c,d}+\xi_{c,}{}^d{}_{,d}) = 0\ , 
\ee 
and the equations
\be 
\lambda \xi^a_{,a} n_c + \mu(\xi^d_{,c}+\xi_{c,}{}^d) n_d = 0\ ,
\ee 
that must hold on the boundary of the body. Since the physical space is flat and three-dimensional, we can easily generalize these results to arbitrary spatial coordinates by replacing partial derivatives with covariant derivatives. The bulk equation becomes
\be
\rho_0 \Ddot \xi^c = \lambda \nabla^c \nabla_a \xi^a + \mu (\nabla_d \nabla^c \xi^d + \nabla_d \nabla^d \xi^c) = 0 \ ,
\ee 
which simplifies to
\be
\Ddot \xi^c = \bigg(\frac{\lambda+\mu}{\rho_0 }\bigg) \nabla^c \nabla_a \xi^a +  \frac{\mu}{\rho_0} \nabla_d \nabla^d \xi^c = 0 \ .\label{eq:bulk}
\ee 
The boundary equation in arbitrary coordinates is 
\be 
\lambda \nabla_a\xi^a n^c + \mu(\nabla^c\xi^d+\nabla^d\xi^c) n_d = 0 \  . \label{eq:bc}
\ee 

The nonrelativistic normal modes of vibration of a solid elastic sphere were first described in a classic paper by Horace Lamb \cite{Lamb1881} in 1881.  See also the later treatise by Love \cite{love1892treatise}.  A modern presentation is given in Thorne and Blandford \cite{Thorne:2017} (exercise 12.12). These normal modes can be separated into two classes, the spheroidal and torsional modes. In this paper, we focus on the spheroidal modes. The subset of the spheroidal modes with $\ell=0$ are called the radial modes. Spherical coordinates, $x^a=\{r,\theta,\phi\}$ are used to simplify the problem. 

We assume a harmonic time dependence.  From \cite{Thorne:2017}, the radial displacement field that satisfies
the bulk Eq.~(\ref{eq:bulk}) is
\be\label{eq:radialmode}
    \vec\xi_n(t,r) = A_n j'_0(\omega_n r/C_L) \, \hat r \, \cos(\omega_n t + \phi_n) \ ,
\ee
where $A_n$ is the amplitude, $\phi_n$ is the phase, and $\omega_n$ is the angular frequency. (At this point, the subscript $n$ is undefined, but will refer subsequently to the discrete modes once the surface boundary condition is imposed and 
the resulting eigenvalue problem is solved.) The spherical Bessel 
functions are denoted by $j_\ell(x)$, with $j'_\ell(x) \equiv \partial j_\ell(x)/\partial x$. The constant $C_L \equiv \sqrt{(\lambda + 2\mu)/\rho_0}$ is the longitudinal sound speed. 

For $\ell > 0$, the general displacement solution satisfying the bulk Eq.~(\ref{eq:bulk}) is \cite{Thorne:2017}
\be 
 \vec{\xi}_{n\ell m}(t,r,\theta,\phi) =  A_{n \ell m}\vec{\Xi}_{n \ell m}(r,\theta,\phi) \cos(\omega_{n \ell}t   + \phi_{n \ell m}) \ ,\label{eq:nmfull}
\ee 
with amplitude $A_{n \ell m}$, phase $\phi_{n\ell m}$ and angular frequency $\omega_{n\ell}$. (Again values for discrete $n$ are yet to be determined.) The vector field $\vec{\Xi}_{n \ell m}$ is given by
\begin{align}    
\vec{\Xi}_{n \ell m} (r,\theta,\phi)&= f_{n  \ell }(r) Y_{ \ell m} {\hat r} \nono\\
&+ g_{n  \ell }(r) \left[ \frac{\pd Y_{ \ell m}}{\pd \theta} {\hat \theta} + \frac{1}{\sin\theta}\frac{\pd Y_{ \ell m}}{\pd \phi} {\hat \phi} \right]\ , \label{eq:nm}
\end{align} 
where $Y_{ \ell m}$ are the real spherical harmonics defined by
\begin{widetext}    
\be 
Y_{\ell m} = \begin{cases}
 \left(-1\right)^m\sqrt{2} \sqrt{\dfrac{2\ell+1}{4\pi}\dfrac{(\ell-|m|)!}{(\ell+|m|)!}} \;
 P_\ell^{|m|}(\cos \theta) \ \sin( |m|\phi ) \ ,
 &\text{if } m<0 \ ,
\\[4pt]
 \sqrt{\dfrac{ 2\ell+1}{4\pi}} \ P_\ell^m(\cos \theta) \ ,
 & \text{if } m=0 \ ,
\\[4pt]
 \left(-1\right)^m\sqrt{2} \sqrt{\dfrac{2\ell+1}{4\pi}\dfrac{(\ell-m)!}{(\ell+m)!}} \;
 P_\ell^m(\cos \theta) \ \cos( m\phi ) \ ,
 & \text{if } m>0 \ ,
\end{cases}
\ee 
\end{widetext}
and $P_\ell^m$ are the associated Legendre functions. The functions, $f_{n \ell }(r)$ and $g_{n \ell }(r)$ are
\begin{align}
f_{n  \ell }(r) &= \frac{\alpha_{n\ell}}{k_{Ln  \ell }}  {j'_\ell(k_{Ln  \ell }r)}{} + \frac{\beta_{n\ell}}{k_{Tn  \ell }}l(l+1) \frac{j_\ell(k_{Tn  \ell }r)}{k_{Tn  \ell }r} \label{eqn:fr}\ , \\
g_{n  \ell }(r) &=  \frac{\alpha_{n\ell}}{k_{Ln\ell}} \frac{j_\ell(k_{Ln  \ell }r)}{k_{Ln\ell}r} \nono\\
&\qq\q+ \frac{\beta_{n\ell}}{k_{Tn  \ell } r} \bigg[ \frac{j_\ell(k_{Tn  \ell } r)}{k_{Tn  \ell }}  + r j'_\ell(k_{Tn  \ell }r) \bigg]\ , \label{eqn:gr}
\end{align}
where, again, $j_\ell(x)$ are the spherical Bessel functions and
\be
k_{Ln\ell} \equiv \frac{\omega_{n\ell}}{C_L} \ , \q
k_{Tn\ell} \equiv \frac{\omega_{n\ell}}{C_T} \ .    
\ee
$C_L$ and $C_T$ are the longitudinal and transverse sound speeds:
\be
C_L = \sqrt{\frac{\lambda + 2 \mu}{\rho_0}} \ , \q
C_T = \sqrt{\frac{ \mu}{\rho_0}} \ . \label{eq:CLCT}   
\ee
The constants $\alpha_{n\ell}$ and $\beta_{nl}$ that appear in the equations for $f_{n\ell}(r)$ and $g_{n\ell}(r)$ determine the weights of the longitudinal and transverse parts of the displacement, with their ratio to be determined when the eigenvalue problem is solved.

These solutions to the bulk motion equation are now subjected to the boundary condition (\ref{eq:bc}), which results in the aforementioned eigenvalue problem.  The eigenvalue problem has an infinite discrete set of solutions, or modes, each marked by an integer $n$.  For each unique spherical harmonic order, these modes differ in their radial dependence and are successively higher frequency overtones.

\begin{figure}
\centering
\includegraphics{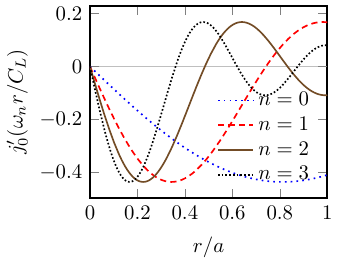}
\caption{\label{fig:fn0} Radial dependence of the  $\ell = 0$ modes, including the $n=0$ fundamental and the first three overtones for $C_L/C_T=\sqrt{3}$. The displacement is zero at the origin for all $n$ values. For the radial  modes, the mode number $n$ coincides with the number of nodes (places of zero displacement) along the radial direction. The maximum displacement does not occur necessarily at the surface.} 
\end{figure}

Let $a$ denote the undeformed radius of the sphere, and $\hat n = \hat r$ denote the unit normal to the boundary. Inserting the   $\ell=0$ radial solution (\ref{eq:radialmode}) evaluated at the surface $r=a$ into the boundary equation (\ref{eq:bc}) results in the following relation:
\be 
\frac{\tan{(\omega_n a/C_L)}}{\omega_n a/C_L} = \frac{4}{4-(\omega_n a/C_T)^2}\ .
\ee
The roots can be obtained numerically for the mode 
frequencies $\omega_n$. The first root corresponds to the first value of $n$ and so on. For example, choosing $C_L/C_T=\sqrt{3}$ we find the solutions for $\omega_{n}a/(\pi C_L)\equiv k_{Ln0}a/\pi$ for $n=0,1,2,3$ shown in Table \ref{tab:kLnl}. Using these solutions, the radial dependence for the $\ell=0$  modes, given by $j'_0(\omega_n r/C_L)$, is plotted in Fig.~\ref{fig:fn0}.

{\renewcommand{\arraystretch}{1.45}
\begin{table*}[tb!]
	\caption{Numerical solutions for $k_{Ln\ell}a/\pi$ for $C_L/C_T=\sqrt{3}$ satisfying the boundary conditions for the first four $\ell$ and $n$ values of the normal modes of oscillation. The values of $k_{Ln\ell}a/\pi$ increase with increasing $n$ number.}
	\label{tab:kLnl}
	\centering
	\begin{tabular*}{\textwidth}{c @{\extracolsep{\fill}} c c c c c }
		\hline
		\hline
		\multicolumn{2}{c }{\quad $\ell$} & $k_{L0\ell}a/\pi$
		& $k_{L1\ell}a/\pi$ & $k_{L2\ell}a/\pi$
		& $k_{L3\ell}a/\pi$ 
		\\
		\hline
		\multicolumn{2}{c}{\quad $0$} & 0.81596643669775 & 1.92853458475813 & 2.95387153514092 & 3.96577216329668
		\\
		\multicolumn{2}{c}{\quad $1$} & 0.62934739815975 & 1.24440286338649 & 1.42338683343041 & 1.96556466385947
		\\
		\multicolumn{2}{c}{\quad $2$} &0.48514540434785 & 0.89412183542721 & 1.53070871073100 & 1.79736223921180
		\\
		\multicolumn{2}{c}{\quad $3$} & 0.71972992130588 & 1.18616009042197 & 1.78353164657311 & 2.15894591358743 
		\\
		\hline
		\hline
	\end{tabular*}
\end{table*}
}

\begin{figure*}
\centering
\includegraphics{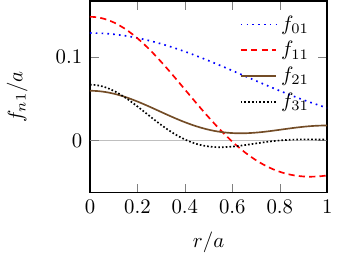}
\includegraphics{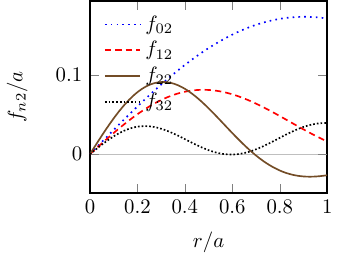}
\includegraphics{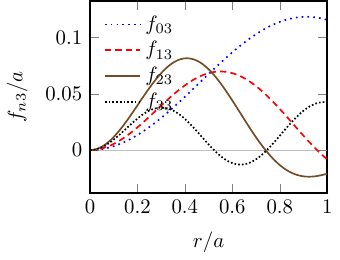}
\includegraphics{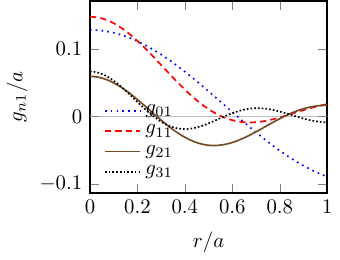}
\includegraphics{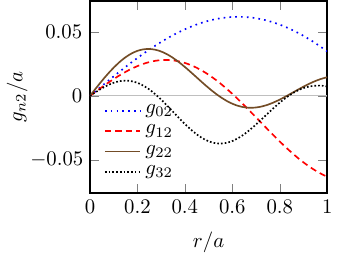}
\includegraphics{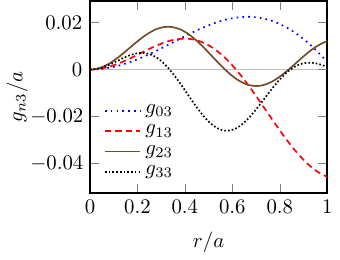}
\caption{\label{fig:fandg} Radial dependence of the functions $f_{n\ell}(r)$ and $g_{n\ell}(r)$ for $C_L/C_T=\sqrt{3}$ and for the first four values of $n$. The top panels show $f_{n\ell }(r)$ for $\ell=1, 2, 3$ (from left to right) and the bottom panels show $g_{n\ell}(r)$. For the $\ell=2$ and $\ell=3$ modes, $\vec{\Xi}_{n\ell m}=0$ at the origin as $f_{n\ell}(r)$ and $g_{n\ell}(r)$ both vanish. This is not true for the $\ell=1$ mode.}  
\end{figure*}

For $\ell > 0$, inserting the bulk displacement solution (\ref{eq:nmfull}) evaluated at the surface $r=a$ into the boundary equation (\ref{eq:bc}) results in two equations,

\begin{align}
\alpha_{n\ell}\big[2 j''_\ell(k_{Ln\ell}a) - ((k_{Tn\ell}/k_{Ln\ell})^2-2))j_\ell(k_{Ln\ell}a)\big] \nono \\
+ \beta_{n\ell}\big[2\ell(\ell+1)f_1(k_{Tn\ell}a)\big] &= 0 \ , \label{eq:bcmodes1} \\
\alpha_{n\ell}\big[2f_1(k_{Ln\ell}a)\big] \qq\qq\qq\qq\qq\qq\q \\
+ \beta_{n\ell}\big[j''_\ell(k_{Tn\ell}a)+(\ell(\ell+1)-2)f_0(k_{Tn\ell}a)\big] &= 0 \ , \nono
\end{align}
where $f_0(x)\equiv j_\ell(x)/x^2$ and $f_1(x)\equiv\pd(j_\ell(x)/x)/\pd x$. The simultaneous linear equations for $\alpha_{n\ell}$ and $\beta_{n\ell}$ have a solution if the determinant is zero,
\begin{align}
&\big[2 j''_\ell(k_{Ln\ell}a) - ((k_{Tn\ell}/k_{Ln\ell})^2-2))j_\ell(k_{Ln\ell}a)\big]\nono\\
&\qq\big[j''_\ell(k_{Tn\ell}a)+(\ell(\ell+1)-2)f_0(k_{Tn\ell}a)\big] \nono \\
&- \big[2f_1(k_{Ln\ell}a)\big]\big[2\ell(\ell+1)f_1(k_{Tn\ell}a)\big]=0 \ . \label{eq:bcdet}    
\end{align}  
Equation (\ref{eq:bcdet}) can be expressed in terms of $k_{Ln\ell}$ and the roots can be obtained numerically. For example, for $C_L/C_T=\sqrt{3}$, we find the solutions for $\ell=1,2,3$ and $n=0,1,2,3$ shown in Table \ref{tab:kLnl}. Inserting these solutions in Eq.~(\ref{eq:bcmodes1}) gives the ratio of the longitudinal to the transverse parts shown in Table \ref{tab:alphaoverbetanl}. Using these solutions, the dependence of the functions, $f_{n\ell}(r)$ and $g_{n\ell}(r)$, on $r$ for $\ell=1,2,3$, and $n=0,1,2,3$ is plotted in Fig.~\ref{fig:fandg}.

{\renewcommand{\arraystretch}{1.45}
\begin{table*}[tb!]
	\caption{Numerical solutions for the ratio of the longitudinal to the transverse parts, $\alpha_{n\ell}/\beta_{n\ell}$, for $C_L/C_T=\sqrt{3}$ for $\ell=1,2,3$ and first four $n$ values of the normal modes of oscillation. The ratios $\alpha_{n\ell}/\beta_{n\ell}$ increase with increasing $\ell$ number.}
	\label{tab:alphaoverbetanl}
	\centering
	\begin{tabular*}{\textwidth}{c @{\extracolsep{\fill}} c c c c c }
		\hline
		\hline
		\multicolumn{2}{c }{\quad $\ell$} & $\alpha_{0\ell}/\beta_{0\ell}$
		& $\alpha_{1\ell}/\beta_{1\ell}$ & $\alpha_{2\ell}/\beta_{2\ell}$
		& $\alpha_{3\ell}/\beta_{3\ell}$ 
		\\
		\hline		
		\multicolumn{2}{c}{\quad $1$} & -0.39334285456883 & 0.57828661556718 & -0.35961617280979 & 0.07851036440781 
		\\
		\multicolumn{2}{c}{\quad $2$} & -0.68808506569504 & -0.95183672540982 & 0.55915283283423 & -1.16213124001178 
		\\
		\multicolumn{2}{c}{\quad $3$} & -1.56275090908497 & -1.65898880431533 & 0.68019269841200 & -1.71401134238877	
		\\
		\hline
		\hline
	\end{tabular*}
\end{table*}
}

\section{Numerical tests}
\label{sec:codeval}

In this section, we use the analytical solutions for an elastic sphere in Sec~\ref{sec:testmodels} to validate the numerical method presented in Sec~\ref{sec:nummethod} (in the nonrelativistic limit) and find its convergence rate.

The numerical method is fully relativistic and is based on nonlinear elasticity. We set the metric equal to the Minkowski metric and choose the analytical solution to be a sum of $\ell=2$ and $\ell=3$ modes with amplitudes $A_{020}$ and $A_{031}$ and phase difference $\phi_{031}-\phi_{020}=\pi/2$:
\begin{align}
\vec{\xi}_{\mathrm{analytic}}(t,r,\theta,\phi) &= \vec{\xi}_{020}(t,r,\theta,\phi) + \vec{\xi}_{031}(t,r,\theta,\phi) \nono \\
 &= A_{020}\vec{\Xi}_{020}(r,\theta,\phi) \cos(\omega_{02}t  + \phi_{020}) \nono \\
 & + A_{031}\vec{\Xi}_{031}(r,\theta,\phi) \cos(\omega_{03}t  + \phi_{031})\ . \label{eq:xitest} 
\end{align} 
We select the material properties of the sphere such that $C_L/C_T=\sqrt{3}$. The amplitudes $A_{020}$ and $A_{031}$ 
are small compared to $a$, and the sound speeds are small compared 
to the speed of light. 

We use Matlab's \cite{pdetool} mesh generation algorithm to generate a linear tetrahedral mesh for a sphere of radius $0.5\,\mathrm{m}$. As the mesh is refined, the total volume of tetrahedral elements converges to $V_\mathrm{conv}$ and we find the converged radius using the converged volume, $a_{\mathrm{conv}}= \sqrt[3]{3 V_{\mathrm{conv}}/4\pi} \approx 0.49881\,\mathrm{m}$. We use $a_\mathrm{conv}$ as the undeformed radius of the sphere in computing the analytical solution. 

We set $X^a$ for all nodes at the initial time step such that their displacement from their relaxed value is equal to Eq.~(\ref{eq:xitest}) evaluated at $t=0$.  We also set $\dot X^a$ equal to the time derivative of Eq.~(\ref{eq:xitest}) evaluated at $t=0$. We numerically evolve the coordinates and velocities in time.

The relativistic terms in the elastic body action are of order $v^2/c^2$  and higher, where $v^2=\dot X^a \dot X_a$ and $c$ is the speed of light. The nonlinear elasticity terms  in the action are of order $(\xi^a_{,i})^3$ and higher. After obtaining the numerical solution we ensured that the discrepancy between the numerical and analytical solution is not due to relativistic and nonlinear elasticity effects by computing $\max(v^2/c^2)$ and $\max(|X^a_{,i}-X^a_{R,i}|)$ using the numerical solution. We found that $\max(v^2/c^2) \approx 10^{-27}$ and $\max(|X^a_{,i}-X^a_{R,i}|) \approx 10^{-8}$, which makes $\max(|X^a_{,i}-X^a_{R,i}|^3)$ about 16 orders of magnitude smaller than $\max(|X^a_{,i}-X^a_{R,i}|)$. 

\begin{figure*}
\centering
\includegraphics{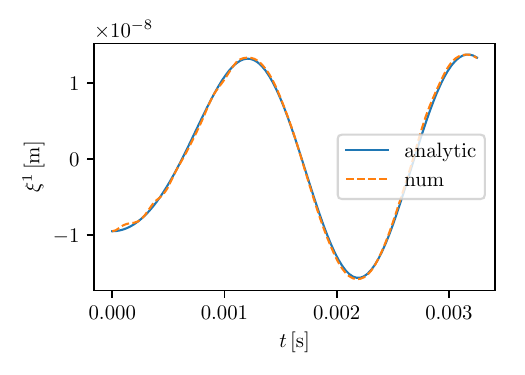}
\includegraphics{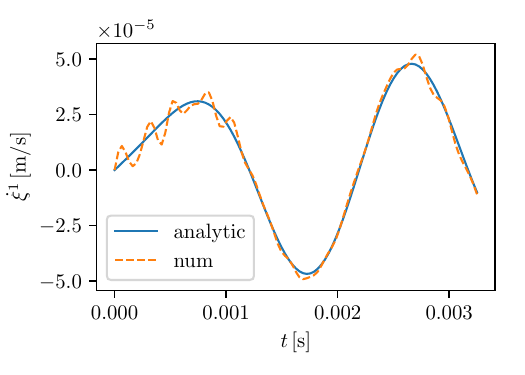}
\includegraphics{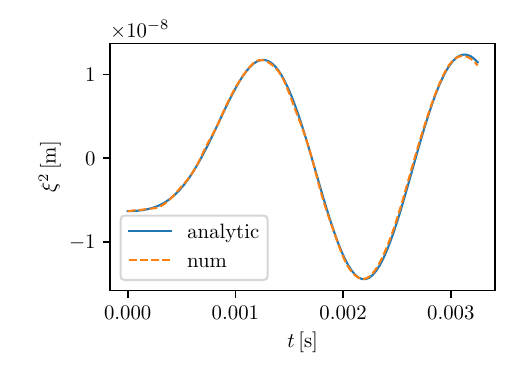}
\includegraphics{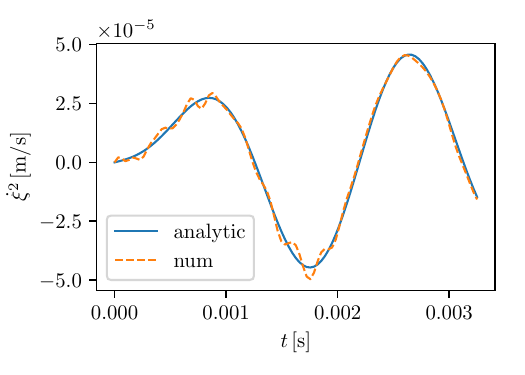}
\includegraphics{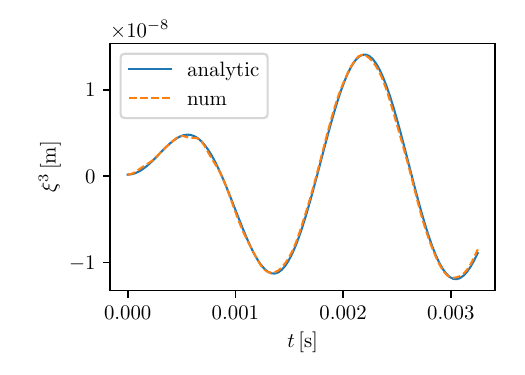}
\includegraphics{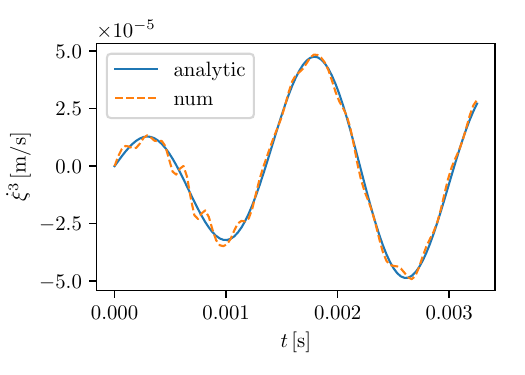}
\caption{ Analytical and numerical displacement and velocity for the node  with matter space coordinates $\zeta^i=(0.1522, 0.2636, -0.3967 )$ as a function of time for the mesh refinement with $h_{\mathrm{max}} = a/8$. The panels on the left show the displacement in three directions and the panels on the right show the velocity. The analytical solution is the sum of $\ell=2$ and $\ell=3$ modes and the material properties of  the sphere are set such that $C_L/C_T=\sqrt{3}.$ For the mesh refinement with $h_{\mathrm{max}} = a/8$, the discretization consists of 3364 nodes and 17403 elements.
\label{fig:flattesth8}}
\end{figure*}

\begin{figure*}
\centering
\includegraphics{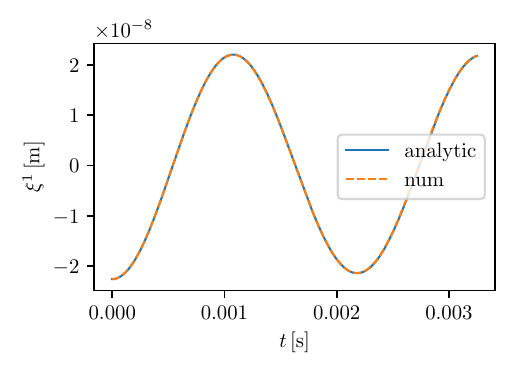}
\includegraphics{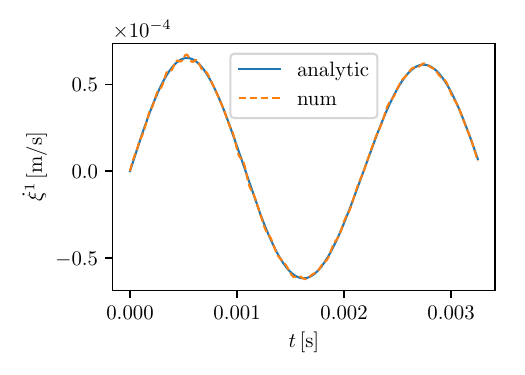}
\includegraphics{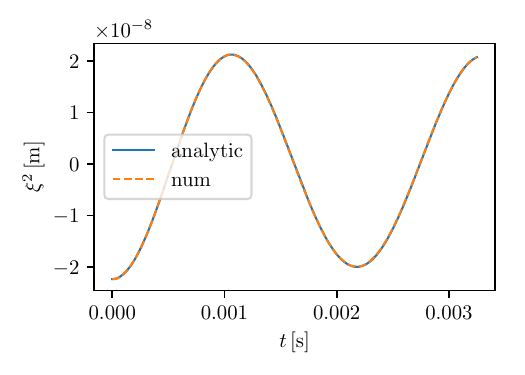}
\includegraphics{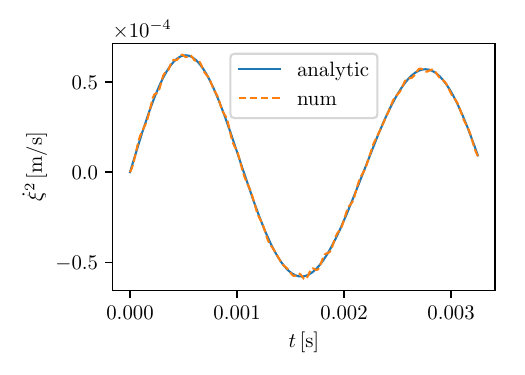}
\includegraphics{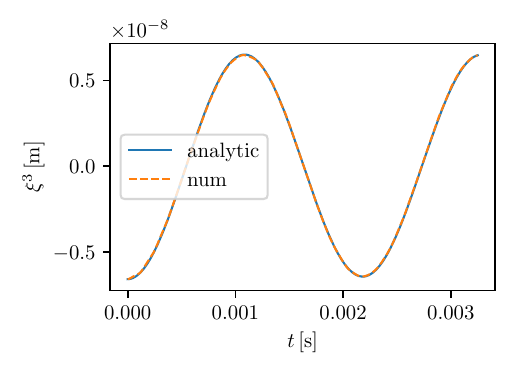}
\includegraphics{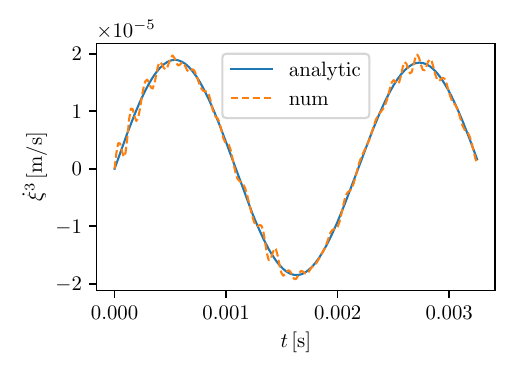}
\caption{ Analytical and numerical displacement and velocity for the node  with matter space coordinates $\zeta^i=(0.1779, 0.4262, 0.1913)$ as a function of time for the mesh refinement with $h_{\mathrm{max}} = a/16$. The panels on the left show the displacement in three directions and the panels on the right show the velocity. The analytical solution is the sum of $\ell=2$ and $\ell=3$ modes and the material properties of sphere are set such that $C_L/C_T=\sqrt{3}.$ For the mesh refinement with $h_{\mathrm{max}} = a/16$, the discretization consists of 25417 nodes and 141001 elements.
\label{fig:flattesth16}}
\end{figure*}

We use four mesh refinements with $h_{\mathrm{max}} = \{a/4, a/8, a/16, a/32\}$, where $h_\mathrm{max}$ is the maximum edge length of the tetrahedral elements. Figure \ref{fig:flattesth8} shows the analytical and numerical displacement and velocity of the node 
$\zeta^i=(0.1522, 0.2636, -0.3967 )$ as a function of time, for the mesh refinement with $h_{\mathrm{max}} = a/8$. Figure \ref{fig:flattesth16} shows the displacement and velocity for the node $\zeta^i=(0.1779, 0.4262, 0.1913)$ with $h_{\mathrm{max}} = a/16$. (The matter space coordinates are Cartesian with metric $\epsilon_{ij} = \delta_{ij}$. The coordinate values are reported  to four decimal places for brevity.) 

\begin{figure*}
\centering
\includegraphics{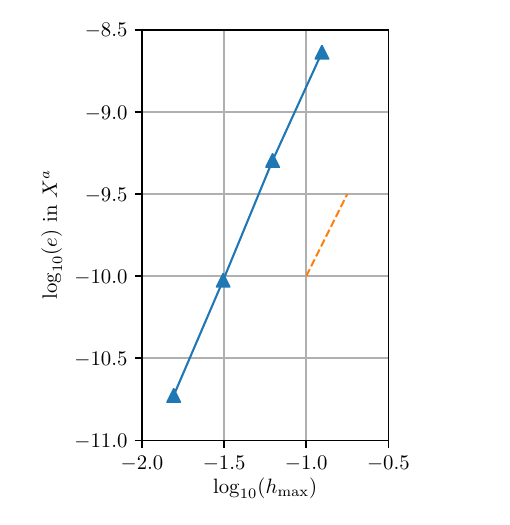}
\includegraphics{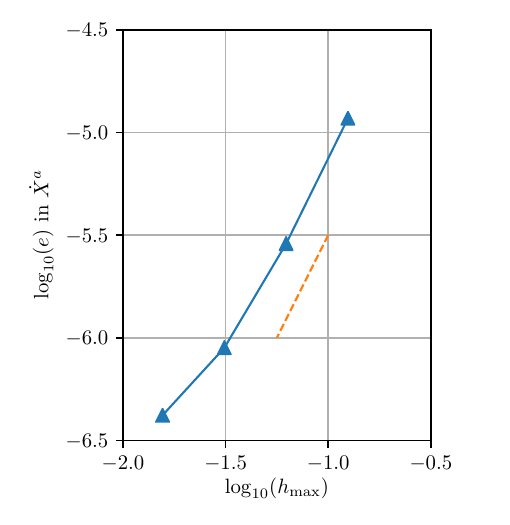}
\caption{Convergence of the L2-norm of the error. The L2-norm of the error at the end of the numerical evolution is plotted against the maximum edge length of tetrahedral elements. The left panel shows the L2-norm of the error in the coordinates and the right panel shows the L2-norm of the error in the velocities. The orange dashed line has slope equal to two. The algorithm displays second order convergence.
\label{fig:conv}}
\end{figure*}

We compute the L2-norm of the error in the coordinates using
\be 
e = \frac{\sqrt{\sum_{n}^{N_\mathrm{total}} (X_n^{a,\mathrm{num}}-X_n^{a,\mathrm{analytic}})(X^{\mathrm{num}}_{a\,n}-X^{\mathrm{analytic}}_{a\,n})}}{N_{\mathrm{total}}}\ ,
\ee 
and similarly for the velocities. Figure \ref{fig:conv} is the log-log plot of the L2-norm of the errors in the coordinates and velocities at the last time step as functions of $h_\mathrm{max}$. The numerical method displays second order convergence.

\section{Conclusions}
\label{sec:summary}

We have presented a second-order-convergent finite element numerical scheme for modeling extended bodies in curved spacetime using elasticity theory in general relativity. Finite elements allow a Lagrangian approach to the elastic body and provide a free surface boundary condition when formulating the numerical method.  The equations of motion for the body are obtained as coupled ODEs by taking a novel approach of spatially discretizing the action. The resulting Euler-Lagrange equations are explicitly integrated in time with fourth--order Runge--Kutta, subject to a Courant condition on the time step. The numerical method can be used for bodies of any shape described by any hyperelastic potential energy function, moving through any spacetime. 

Reducing to a linearized action for the hyperelastic body in the nonrelativistic limit, we reproduced the classic solutions \cite{Lamb1881}\cite{Thorne:2017} for radial and nonradial normal mode oscillations of an elastic sphere.  These modes were then used to test the numerical code in the linearized, nonrelativistic limit. By ensuring that relativistic and nonlinear contributions are negligible, our numerical results show second-order convergence to the analytical solutions. 

In a paper to follow shortly, we will apply our numerical framework to model the motion and internal dynamics of a hyperelastic sphere during tidal encounters with a Schwarzschild black hole along a quasi-parabolic orbit. Beyond that, the method presented in this paper will allow a host of investigations to be carried out on extended body interactions with spacetime curvature, including MPD spin-curvature effects on rapidly-rotating bodies and effects of higher multipole moments.  Encounters could be generalized to scattering with Kerr black holes.  Furthermore, the technique could be extended to include gravitational perturbations and radiation reaction effects on the finite-sized mass.

\begin{acknowledgments}
We acknowledge the computing resources provided by North Carolina State University High Performance Computing Services Core Facility (RRID\:SCR\_022168). We also thank Lisa L. Lowe for her assistance with porting and optimization. C.R.E.~was partially supported by NSF Grant No.~PHY-2110335 to the University of North Carolina--Chapel Hill.
\end{acknowledgments}

\bibliography{ref}
\vfill\eject

\end{document}